\documentclass[]{spie}  

\usepackage{amsmath,amsfonts,amssymb}
\usepackage{graphicx}
\usepackage[colorlinks=true, allcolors=blue]{hyperref}

\title{The Keck Planet Imager and Characterizer: Phase I fiber injection unit early performance and commissioning}

\author[a]{Evan C. Morris}
\author[b]{Jason J. Wang}
\author[b]{Jean-Baptiste Ruffio}
\author[b,d]{Jacques-Robert Delorme}
\author[b]{Jacklyn Pezzato}
\author[d]{Charlotte Z. Bond}
\author[b,c]{Dimitri Mawet}
\author[a]{Andrew J. Skemer}
\affil[a]{U.C. Santa Cruz, 1156 High Street, Santa Cruz, Ca 95064, USA.}
\affil[b]{Department of Astronomy, California Institute of Technology, Pasadena, CA 91125, USA}
\affil[c]{Jet Propulsion Laboratory, California Institute of Technology, 4800 Oak Grove Dr., Pasadena, CA 91109, USA}
\affil[d]{W. M. Keck Observatory, 65-1120 Mamalahoa Highway., Kamuela, HI 96743, USA.}

\authorinfo{Further author information: (Send correspondence to Evan Morris)\\Evan Morris: E-mail:  ecmorris@ucsc.edu}

\begin{document} 
\maketitle

\begin{abstract}
The Keck Planet Imager and Characterizer (KPIC) is an upgrade to the Keck II adaptive optics system and instrument suite with the goal of improving direct imaging and high-resolution spectroscopic characterization capabilities for giant exoplanets. KPIC Phase I includes a fiber injection unit (FIU) downstream of a new pyramid wavefront sensor, coupling planet light to a single mode fiber fed into NIRSPEC, Keck's high-resolution infrared spectrograph. This enables high-dispersion spectroscopy (HDS) of directly imaged exoplanets at smaller separation and higher contrast, improving our spectral characterization capabilities for these objects. Here, we report performance results from the KPIC Phase I FIU commissioning, including analysis of throughput, stability, and sensitivity of the instrument. 
\end{abstract}


\section{INTRODUCTION}
\label{sec:intro}  

High-dispersion spectroscopy is a powerful tool in exoplanet atmospheric characterization~\cite{Brogi2019,Snellen2014,Snellen2015_A&A,Wang2017_AJ,Mawet2017_APJ}. By resolving molecular lines, it allows more accurate analysis of molecular abundances, planetary spin measurements, and radial velocity measurements, and the potential for Doppler imaging of atmospheric features. These characterization techniques were limited to wide orbit companions using the previous generation of high-resolution spectrographs which were not designed to address the challenges of high-contrast. 

The Keck Planet Imager and Characterizer (KPIC)\cite{Mawet2016_SPIE,Mawet2018} is optimized for obtaining these high-resolution spectra of high contrast planets, increasing sensitivity and precision when detecting and characterizing these atmospheres. KPIC is an upgrade to the Keck II adaptive optics system, paired with a fiber injection unit (FIU) to NIRSPEC~\cite{McLean1998_SPIE,Martin2018_SPIE}, Keck’s existing high-resolution near-infrared spectrograph. When injecting light into NIRSPEC using KPIC, light downstream of the infrared pyramid wavefront sensor (PWS)~\cite{Bond2020_SPIE,bond2020-AOI} is coupled into an array of single mode fibers in the FIU, allowing for high-dispersion spectroscopy in K and L band. KPIC uses single mode fibers to isolate light from a planet from that of its host star, with the fibers providing excellent starlight rejection~\cite{jovanovic2017_AO4ELT} and a stable line spread function. 

In order to take advantage of these capabilities, KPIC must be carefully aligned. Small tolerances mean that, in order to maximize coupling into the spectrograph, the planet must be within a small fraction of a diffraction limited point spread function (see Section \ref{sec:offsky}). In this paper, we present an overview of early performance of KPIC’s FIU in commissioning, focusing on throughput, stability, and blind offsetting capabilities.

\section{INSTRUMENT OVERVIEW}
\label{sec:instrument}

KPIC demonstrates new exoplanet characterization technology and techniques. It is deployed at the W. M. Keck Observatory, located downstream of the Keck II adaptive optics system~\cite{Wizinowich2000_PASP}. The project is broken down into multiple phases, with phase I currently deployed and phase II planned for fall 2021\cite{Jovanovic2019_SPIE,pezzato2019-SKP,jovanovic2020-PVC}.

\begin{figure*}[h]
\includegraphics[width=6.5in]{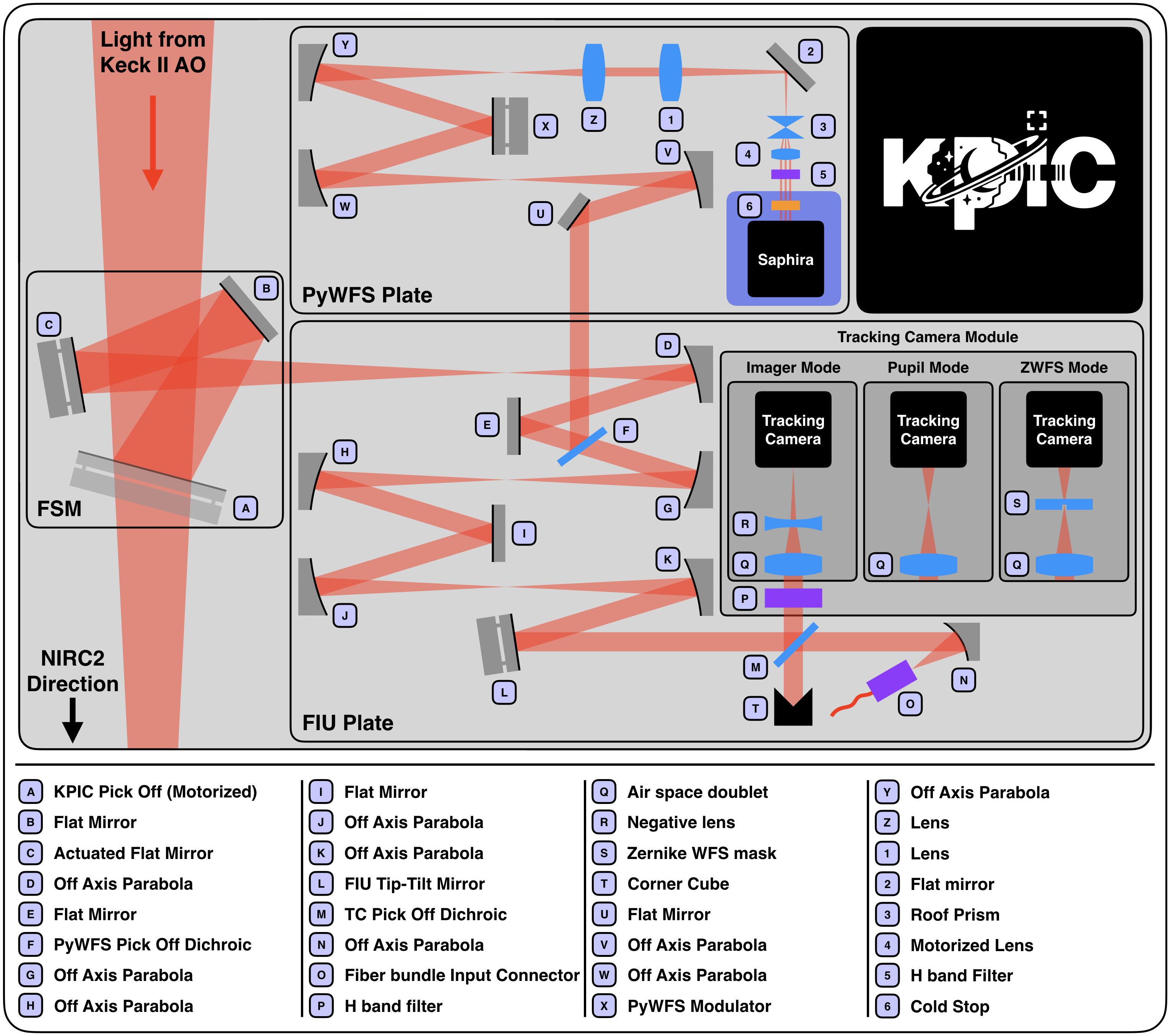}
{\caption{KPIC optical layout.}\label{fig:kpic_layout}}
\end{figure*}

Phase I consists of connecting the infrared pyramid wavefront sensor (PWS)~\cite{Bond2020_SPIE,bond2020-AOI} to the existing infrared spectrograph, NIRSPEC, using single mode fibers. The FIU components of KPIC are an injection module, a fiber bundle, and an extraction module~\cite{Delorme2020}. See Figure \ref{fig:kpic_layout}.

The injection module is used to inject light of high contrast companions, such as exoplanets and brown dwarfs, into a single-mode fiber (SMF) using a tip-tilt mirror (TTM) in the pupil plane of the instrument. A custom dichroic redirects J and H band light to a tracking camera, and transmits K and L band light to the science instrument. The tracking camera and the TTM are used in a closed loop to ensure the alignment of the companion with the fiber. 

The fiber bundle is positioned at the focus and on the optical axis of an off-axis parabola (OAP). It is used to route light from the injection module into the spectrograph, and includes calibration fibers than can be used to inject light in the reverse direction for alignment purposes. The bundle contains five science fibers and two calibration fibers, and is 5 meters long~\cite{Delorme2020}. The number of science fibers was determined by the need for at least one planet fiber, one star fiber, and one background fiber, with additional fibers for redundancy and flexibility.

The fiber extraction unit (FEU) is the final part downstream of KPIC, acting as an interface between the fiber bundle and the existing spectrograph, NIRSPEC. It reshapes and injects the light from all of the fibers into the NIRSPEC slit. 

There is one KPIC specific part within NIRSPEC itself, an additional cold stop added to a NIRSPEC filter wheel. This stop was installed in an effort to increase KPIC phase I throughput, as a replacement for the more general AO stop, which isn’t optimized for the background versus light tradeoff for KPIC specifically, resulting in a roughly 50\% throughput loss.

\section{OBSERVING PROCEDURE}
\label{sec:observing}

Over the course of commissioning KPIC phase I, we have optimized our observing procedure. Before a run, we take calibration data focused on evaluating and updating the calibration and distortion maps for the tracking camera, as well as verifying the fiber locations on that detector. Once on sky, we confirm the distortion solution by imaging a bright binary. For each night, we also take data on an RV standard M-giant star to create a wavelength solution, which can be shifted linearly over the course of the night as needed. For telluric calibration, we use spectra of on the host star or a nearby A0 star, depending on the observation.  

\subsection{Off-sky Calibration}
\label{sec:offsky}
\begin{figure*}[h]
\includegraphics[width=6.5in]{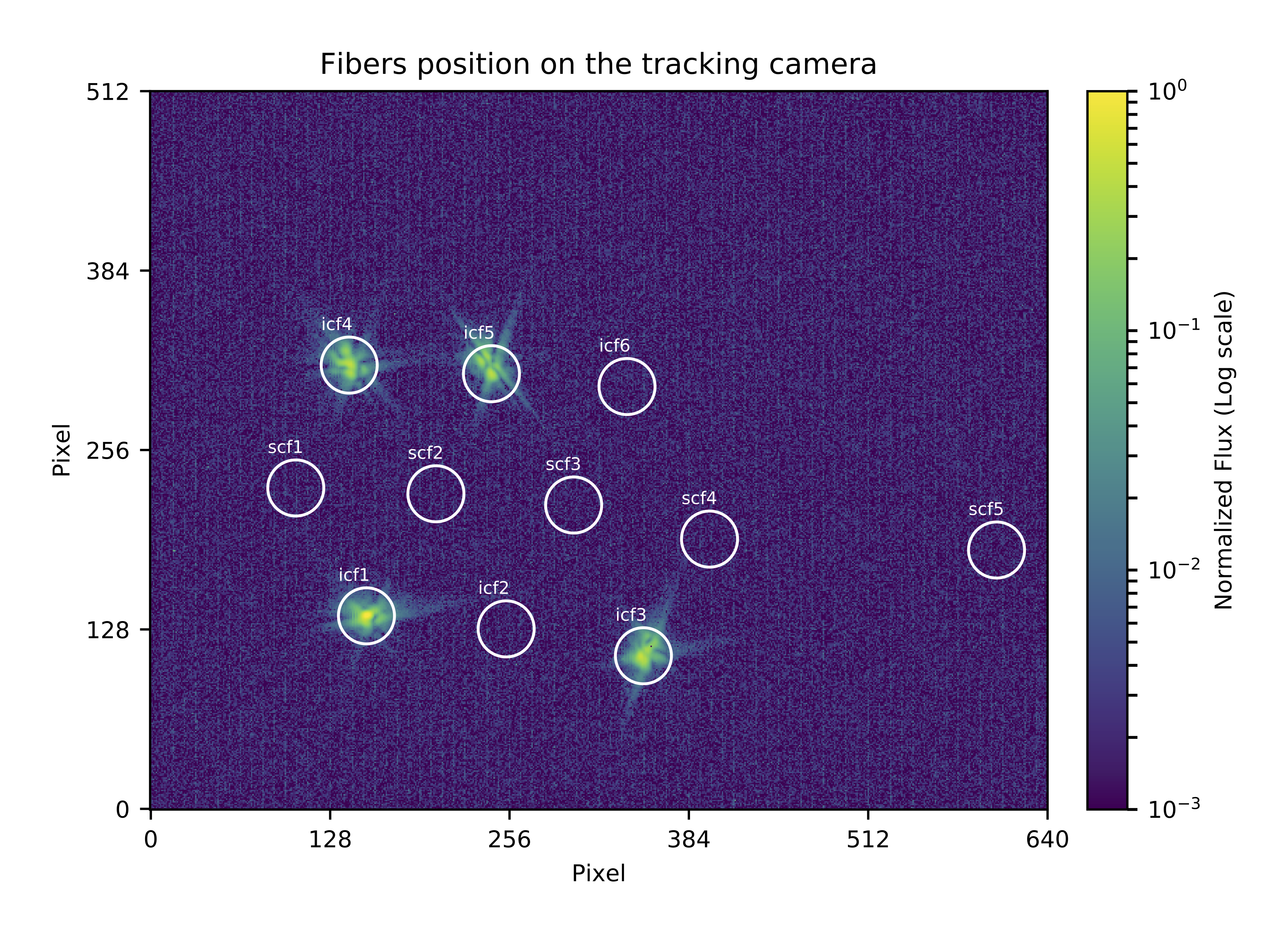}
{\caption{Four calibration fibers used to roughly locate the science fiber locations on the tracking camera.}\label{fig:tracking_cam}}
\end{figure*}

Before each run, we verify the fiber positions on our tracking camera’s detector. This is done through a two step process, a fast but not highly accurate guess, and a more in-depth follow-up optimization. Initially, the fiber positions are unknown, as they are not visible on any sensors. In our first step, light from an infrared laser is injected into four calibration fibers and emitted from the input connector of the bundle, forming an image of the four PSFs on the tracking camera, as seen in Figure \ref{fig:tracking_cam}. Knowing the geometry of the bundle, these positions can be used to roughly locate the science fibers.

\begin{figure*}[h]
\includegraphics[width=6.5in]{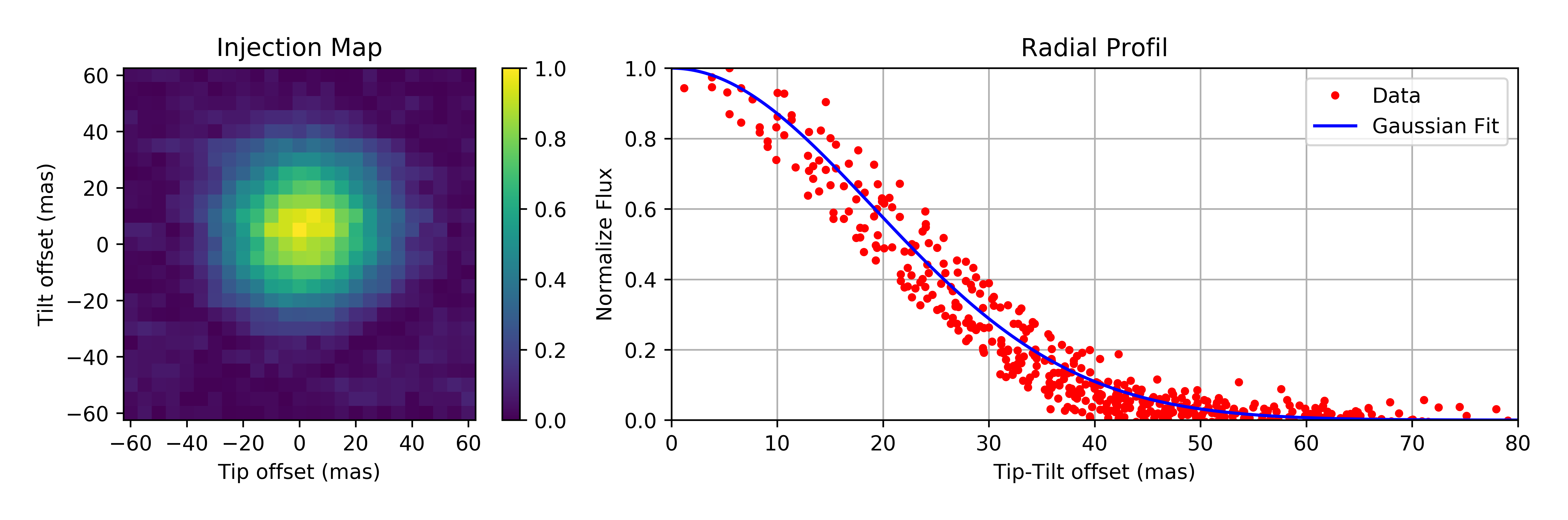}
{\caption{Injection map radial profile showing losses when offset from the fiber position.}\label{fig:injection}}
\end{figure*}

These results can be off by a couple of pixels for a number of reasons, so we use a second check to get a more accurate result. For this step, we directly measure light injected into the fibers using the calibration source of the Keck II AO bench. This creates a PSF on the tracking camera, and we can align this PSF with one of the science fibers using the FIU tip tilt mirror (TTM). We can scan across the expected location of the fiber and record flux on SCAM, the NIRSPEC slit-viewing camera and maximize the transmitted flux. With this method, we can find the optimal TTM position and PSF position on the tracking camera that best couples light from the calibration source into our science fiber. We repeat this measurement for each fiber.

Though this process is time-consuming, we perform it on each night that we are observing. Finding this ideal position as accurately as possible is essential for optimizing throughput. As shown in Figure \ref{fig:injection}, An incorrect offset of one pixel on the tracking camera (8.06 mas) corresponds to a loss of 10\% of throughput, with two pixels (16.12 mas) increasing this loss to 30\%, with  55\% and higher for offsets of three pixels (24.18 mas) and greater. An offset corresponding to the angular resolution of the Keck telescope, 50 mas at 2$\mu$m, would result in a 97\% throughput loss.

For this process, it is critical that the tracking camera itself is well calibrated in terms of distortion, plate scale, and orientation. We use two methods for this, first an off-sky procedure on the day prior to observing. We use NIRC2 as a reference, as its plate scale and distortion is well characterized, imaging the PSF of the calibration source on both the NIRC2 and tracking camera detectors simultaneously. We compute the position of the PSF in several positions to determine the tracking camera plate scale and orientation, the variation of which indicates a distortion solution.

We also perform NIRSPEC specific calibrations in advance, as KPIC uses the spectrograph in a non-standard way. Before observing, we take backgrounds and flats. We can also inject the calibration source into each science fiber individually to locate the full trace on the spectrograph’s detector, which can make identifying fiber traces easier while doing quick data reduction during the run. We perform these calibrations for each instrument configuration we plan to use, though to this point we have also usually used only one configuration per run.

\subsection{On-sky Calibration}
Once we go on sky, we check our off-sky calibrations against data on sky. We use NIRC2 to image a bright binary star system and calculate its separation and position angle, using these along with our off-sky or previously calculated distortion solution to offset from the primary to the secondary on our main science fiber. Following this offset, we check to make sure that the secondary has been moved to the correct fiber position on the tracking camera’s detector. This assures us that our distortion map is accurate and that our subsequent blind offsets to dim planetary companions will be successful. If this check is unsuccessful, we use the same binary star, or a set of such systems to create a new distortion solution. In this case, we tile the binary over the detector using the TTM, measuring the difference in separation and position angle in each location to create a map of these differences. With this method, plate scale and orientation can be determined using an average of the measurements. 

On sky, we repeat our spectrograph calibrations. When observing a bright star, we iterate over the science fibers, taking quick spectra of each to check each fiber’s position through the NIRSPEC slit. This avoids the potential issue of thermal changes in the instrument between its daytime and nighttime uses. 

We also calculate our wavelength solution using a standard on-sky calibration. We find that the wavelength solution is stable within a small margin, so we are able to use one detailed measurement as a baseline for all of our night’s science data. We take spectra of a well-characterized M-giant star, as well as a telluric standard nearby. These M stars have abundant, deep, and well understood lines throughout our K band orders, allowing for more precision than calibrating only using tellurics would allow. 

Before and after observing a target, as well at one hour intervals during a long observation, we observe a telluric standard star either nearby our science target, or use the host star itself if the stellar spectrum is close to featureless. This gives us accurate telluric correction, as well as allowing us to fit the small shift in the wavelength solution that occurs over the course of the night. 

\subsection{Data Acquisition}
\begin{figure*}[h]
\includegraphics[width=6.5in]{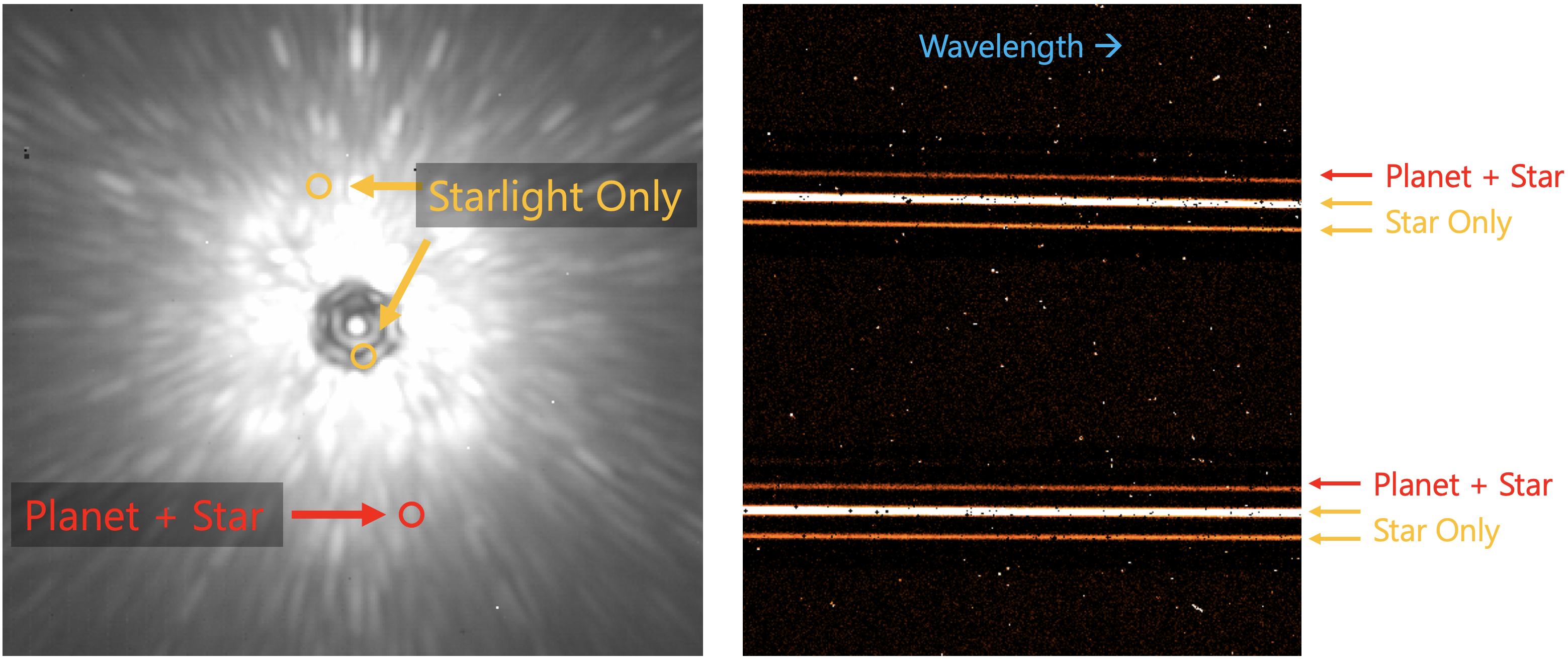}
{\caption{Fiber positions on sky and on the NIRSPEC detector. One fiber is placed on the planet or brown dwarf companion, one fiber is placed near the star, and thee other fibers are on sky and capture varying amount of stellar light.}\label{fig:fiberonsky}}
\end{figure*}

KPIC is optimized to observe high-contrast companions. When beginning a science target observation sequence on a planet or brown dwarf companion, we first align the host star with the selected fiber, and take data on it using at least the primary and secondary science fibers. Figure \ref{fig:fiberonsky} shows the fiber positions on sky and on the spectrograph detector while exposing on the companion. We then offset to the companion and take multiple 600 second exposures We return to the host for calibration purposes and in order to check throughput once every hour, which changes based on AO correction quality, weather, and the target's airmass. While observing the host and the companion, we track using the brighter host, using the TTM in the FIU to keep the target aligned with our fiber by fitting the stellar PSF and comparing its location to where we expect it to be at all times. 

KPIC FIU data is formatted identically to NIRSPEC data taken after its recent upgrade, consisting of a series of 2048x2048 pixel images. Each frame contains 9 spectral orders covering the K band wavelength range (1.94 - 2.48 microns).

\subsection{Data Reduction}
We reduce our data and extract spectra using a custom-built pipeline. Using sets of background frames taken with the same exposure times as our science and calibration data, we create a combined background frame using a mean of each set, find the noise of each pixel using the standard deviation of the frames, and locate bad pixels on the NIRSPEC detector. 
Instead of rectifying orders, we locate each fiber on the detector and extract based on its profile directly. Thanks to the single mode fiber, the line width is very stable over time. In order to identify the traces of the primary and secondary science fibers, we read in all of the data on our calibrator star and take the mean to create a single 2-D frame. We subtract the corresponding background and mask bad pixels. We identify the rough fiber location and fit a 3rd order polynomial to model the full trace on the detector. We low pass filter the data to get the instrumental spectral response of each fiber. 

For our spectral extraction, we load our science frames, subtract the appropriate background, and mask bad pixels. We offset our traces to extract the background. We use the median of these background traces as an additional term to subtract off in the following step, to compensate for imperfect background subtraction. We use polynomial coefficients from measured spectral traces to fix the trace centers, and fix the standard deviation to the median computed in the trace calibration. We fit for the amplitude of the Gaussian and convert amplitude to total flux, weighing each pixel by the noise computed from the background frames and photon noise from the data itself. We take this to be the flux for this slice of the trace, and repeat across the slit.

\begin{figure*}[h]
\includegraphics[width=6.5in]{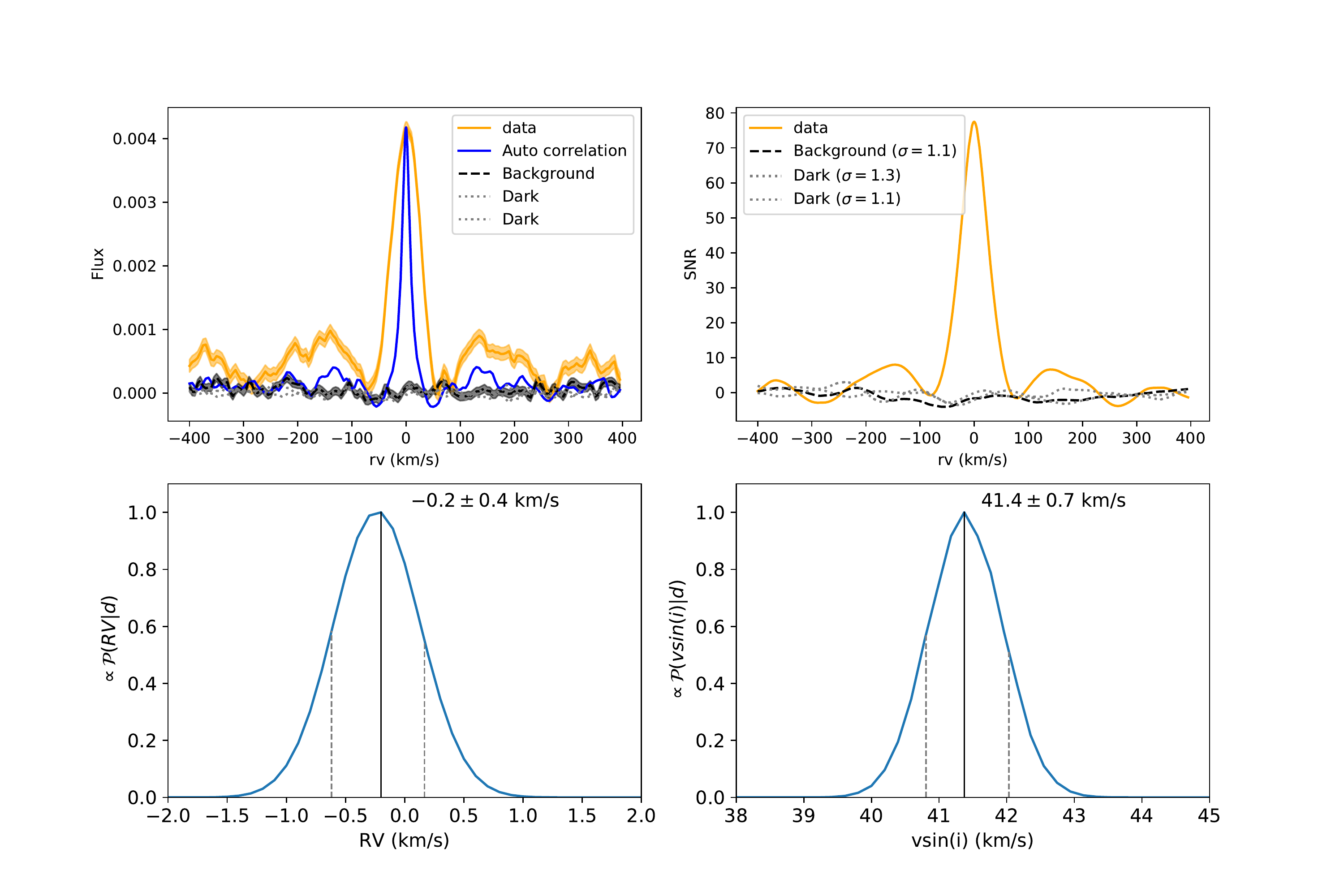}
{\caption{Detection of HR 7672 B using a cross-correlation method. Top left shows estimated flux when radial velocity (RV) is varied and spin is held at zero. Top right shows SNR when assuming best fit spin (vsin(i) = 41.4 km/s). Bottom panels show RV and spin posteriors.}\label{fig:xcorr}}
\end{figure*}

For our wavelength calibration, we use data from a bright M star. We compute expected RV shift of the star using barycentric correction and the systemic RV. We use a PHOENIX stellar model, a telluric model of the atmosphere, and the spectral response of instrument to build a model of the starlight, adjusted for continuum and background. For each order, we fit for wavelength using a spline, first performing a grid search with three nodes for an initial fit, and then running a non-linear simplex optimizer (Nelder-Mead) to identify a best fit. We do this once per night, and use telluric information in our science observations to linearly shift this wavelength solution to fit an earlier or later dataset.

Our data reduction process also includes preliminary measurement of companion radial velocity and spin, as well as checking for detection of expected molecules. We construct a forward model of the companion's spectrum using our extracted stellar spectra, wavelength solution, measured spectral response, and planetary atmosphere models. We vary the radial velocity spin, and flux of the companion when fitting to our data. Figure \ref{fig:xcorr} shows the best fit model when varying over radial velocity, as well as our preliminary spin measurement for HR 7672 B~\cite{Liu2002_ApJ}. More detail on this fitting technique will be included in upcoming papers on our results.

\section{INSTRUMENT PERFORMANCE}
\label{sec:performance}
 
\subsection{Throughput}

Throughput has been the limiting factor determining final performance, and a major driver of instrument and observing procedure changes within KPIC that we have made during commissioning. High-dispersion spectroscopy requires enough photons per spectral channel to detect individual lines. We are background noise limited due to the highly dispersed spectra, making throughput critical. When using single mode fibers to observe directly imaged exoplanets, the total amount of light available is small, so preserving as much of that as possible through the system is essential. 

\begin{figure*}[h]
\includegraphics[width=6.5in]{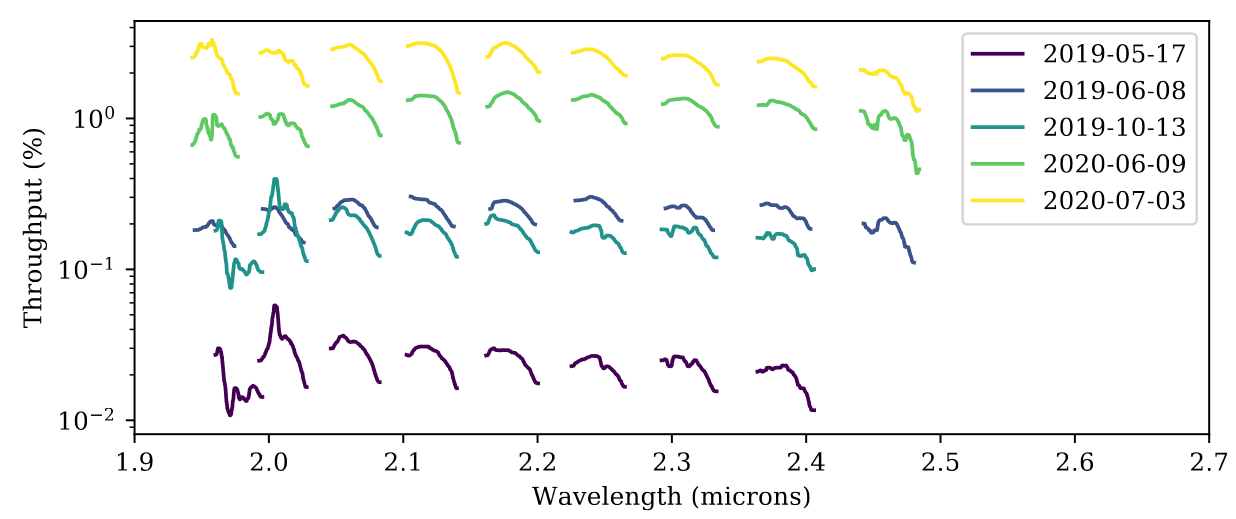}
{\caption{Throughput increased with each change within thee instrument. The first major change relates to improving alignment, and the second relates to the introduction of the FEU stop.}\label{fig:throughput}}
\end{figure*}

Currently, KPIC’s peak throughput in K band is 3.2\%, as measured through the second science fiber on our clearest night of observing. Figure \ref{fig:throughput} represents progress through a year of observations and changes, with measurements taken at the peak of each run. Each change inside the instrument or in our calibration process has brought a major improvement. While factors such as seeing and weather affect this number out of our control, we are consistently able to measure between 1.5 and 2.5\% throughput over the K band in our last few runs. These measurements are consistent with the expected throughput for KPIC phase 1, calculated to be up to 3.4\%, assuming current performance of each optic and allowing for 200 nm RMS of residual aberration~\cite{Delorme2020}. 

Phase 1 observations have focused on the K band, but preliminary throughput measurements have also been made in the L band, with NIRSPEC’s KL filter, reaching 6\% between 3.7 and 3.8 $\mu$m, but at these wavelengths instrument and sky background are also much higher. Future improvements to KPIC in phase 2, including the inclusion of an an atmospheric dispersion compensator (ADC), will allow us to optimize throughput for all wavelengths and see an overall increase. 

\subsection{Blind Offsetting}

In our blind offsets, we are limited by our knowledge of system astrometry, as these objects are generally not able to be imaged on the tracking camera, or even with NIRC2 without a longer sequence of imaging observations. From this information, we find the separation and position angle of the offset, and use the telescope rotator to shift our orientation such that the secondary science fiber is near but not overlapping the star while the primary science fiber is observing the planet.

When evaluating future areas to improve, we see some indication that or ability to detect planets in high contrast settings is more developed than our blind offsetting capabilities. With the HR 8799 system, we detect the close in HR 8799e with more ease than HR 8799b, when we might otherwise expect a quicker detection of b. This suggests that, while our techniques to filter out star light are successful, we need to improve our offsets, though we have not yet quantified this issue.

\subsection{Wavelength Stability}

\begin{figure}[h]
\includegraphics[width=6.5in]{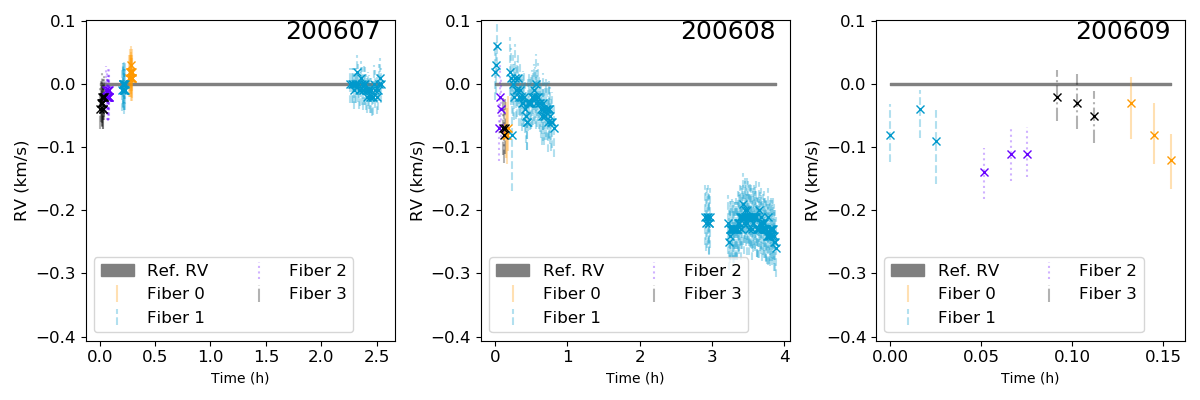}
{\caption{Wavelength solution shift in relation to reference radial velocity during three nights of observations, as measured on four science fibers. On night one, we see that our solution is stable over a 2.5 hour period. On night two, we see a shift due to movement within the instrument.}\label{fig:wavelength}}
\end{figure}

Stability in the wavelength solution is related to a number of factors within KPIC and within NIRSPEC. The wavelength solution changes across the NIRSPEC detector, meaning that KPIC’s consistency in fiber position on the detector has an advantage over traditional NIRSPEC slit observations, in which object position in the slit is less consistent. But even looking at the same detector locations, the solution changes slowly due to thermal shifts. We measure this change to be subpixel over the course of the night, allowing for our use of one detailed wavelength solution using an M star. If needed, this solution can be linearly shifted using tellurics to fit later and earlier data, but due to the stability this is not generally required. This changes when parts of the instrument move, such as the grating, which does not return perfectly to previous settings, so we try not to change any NIRSPEC configuration settings during a run. Figure \ref{fig:wavelength} shows the wavelength solution stability

\subsection{Thermal Background}

For the KPIC FIU, our dominant source of noise is the thermal background. Most of what we see is in the NIRSPEC instrument, both between the fiber bundle and NIRSPEC cold stop and in the cryostat. Single mode fibers are highly efficient at rejecting upstream light. Emissivity of the fibers themselves has not been quantified, but should also be negligible, as KPIC uses short (5 m), narrow (core-diameter of $6.5~\mu$m) fibers with transmission of $>99\%$ in K band. This indicates that sky background and other optics will add very little to our total background. One purpose of the addition of the FEU stop in NIRSPEC was to optimize the signal-to-noise ratio seen at the detector, in terms of minimizing thermal background while maximizing photons from our target. Previously, we used the general AO stop, which cut down the background more than the FEU stop, but also blocked more signal. A main source of background is within NIRSPEC itself, where a known bright optical leak adds thermal noise. NIRSPEC also has a fringing pattern that changes over the course of observations.

\section{FUTURE WORK}
\label{sec:future}
Going forward in KPIC phase I, we are continuing to test and quantify aspects of our performance. These calibration tests will include a more extensive throughput analysis, comparing NIRSPEC data to simultaneous tracking camera data, as well as investigating the impact of atmospheric conditions. We will quantify our long term wavelength stability, line spread function stability, and our error budget for detecting exoplanets. We are also currently improving our post-processing capabilities, in the form of a streamlined and KPIC-optimized pipeline.

Future improvements to KPIC in phase II, including the addition of a 952 element deformable mirror, phase induced amplitude apodization lenses, an atmospheric dispersion compensator, multiple coronagraphs, a Zernike wavefront sensor, and multiple science ports, will further optimize the system throughput and contrast.

\acknowledgments 
This work was supported by the Heising-Simons Foundation through grants \#2019-1312 and \#2015-129. Part of this work was carried out at the Jet Propulsion Laboratory, California Institute of Technology, under contract with the National Aeronautics and Space Administration (NASA). W. M. Keck Observatory is operated as a scientific partnership among the California Institute of Technology, the University of California, and the National Aeronautics and Space Administration (NASA). The Observatory was made possible by the generous financial support of the W. M. Keck Foundation. The authors wish to recognize and acknowledge the very significant cultural role and reverence that the summit of Maunakea has always had within the indigenous Hawaiian community. We are most fortunate to have the opportunity to conduct observations from this mountain.  

\bibliography{main} 
\bibliographystyle{spiebib} 

\end{document}